\documentclass[%
 reprint,
superscriptaddress,
showpacs,preprintnumbers,
 amsmath,amssymb,
 aps,
]{revtex4-1}

\usepackage{dcolumn}
\usepackage{bm}
\usepackage[mathlines]{lineno}

\usepackage{amsmath}
\usepackage{graphicx}
\usepackage{dcolumn}
\usepackage{colortbl}
\usepackage{multirow}
\usepackage{graphicx}
\usepackage{booktabs}
\usepackage{array}
\usepackage{ amssymb }
\usepackage{tabularx}
\usepackage[table,xcdraw]{xcolor}
\usepackage{bm}
\usepackage{hyperref}
\usepackage[mathlines]{lineno}

\begin{document}

\title{Kinetic simulations of fusion ignition with hot-spot ablator mix}

\author{James D. Sadler}
\affiliation{%
Los Alamos National Laboratory, P.O. Box 1663, Los Alamos, New Mexico 87545, USA
}%
\affiliation{%
 Clarendon Laboratory, University of Oxford, Parks Road, Oxford OX1 3PU, UK
}%

\author {Yingchao Lu}
\affiliation{%
Los Alamos National Laboratory, P.O. Box 1663, Los Alamos, New Mexico 87545, USA
}%

\author {Benjamin Spiers}
\affiliation{%
 Clarendon Laboratory, University of Oxford, Parks Road, Oxford OX1 3PU, UK
}%
\author {Marko W. Mayr}
\affiliation{%
 Clarendon Laboratory, University of Oxford, Parks Road, Oxford OX1 3PU, UK
}%
\author {Alex Savin}
\affiliation{%
 Clarendon Laboratory, University of Oxford, Parks Road, Oxford OX1 3PU, UK
}%
\author {Robin H. W. Wang}
\affiliation{%
 Clarendon Laboratory, University of Oxford, Parks Road, Oxford OX1 3PU, UK
}%
\author {Ramy Aboushelbaya}
\affiliation{%
 Clarendon Laboratory, University of Oxford, Parks Road, Oxford OX1 3PU, UK
}%

\author {Kevin Glize}
\affiliation{%
Central Laser Facility, STFC Rutherford Appleton Laboratory, Didcot, OX11 0QX, UK
}%
\author {Robert Bingham}

\affiliation{%
Central Laser Facility, STFC Rutherford Appleton Laboratory, Didcot, OX11 0QX, UK
}%
\affiliation{%
Department of Physics, University of Strathclyde, 107 Rottenrow East, Glasgow, G4 0NG, UK
}%

\author {Hui Li}
\affiliation{%
Los Alamos National Laboratory, P.O. Box 1663, Los Alamos, New Mexico 87545, USA
}%

\author {Kirk A. Flippo}
\affiliation{%
Los Alamos National Laboratory, P.O. Box 1663, Los Alamos, New Mexico 87545, USA
}%

\author {Peter A. Norreys}
\affiliation{%
 Clarendon Laboratory, University of Oxford, Parks Road, Oxford OX1 3PU, UK
}%
\affiliation{%
Central Laser Facility, STFC Rutherford Appleton Laboratory, Didcot, OX11 0QX, UK
}%

\begin{abstract}
Inertial confinement fusion fuel suffers increased X-ray radiation losses when carbon from the capsule ablator mixes into the hot-spot. Here we present one and two-dimensional ion Vlasov-Fokker-Planck simulations that resolve hot-spot self heating in the presence of a localized spike of carbon mix, totalling $1.9\,\%$ of the hot-spot mass. The mix region cools and contracts over tens of picoseconds, increasing its alpha particle stopping power and radiative losses. This makes a localized mix region more severe than an equal amount of uniformly distributed mix. There is also a purely kinetic effect that reduces fusion reactivity by several percent, since faster ions in the tail of the distribution are absorbed by the mix region. Radiative cooling and contraction of the spike induces fluid motion, causing neutron spectrum broadening. This artificially increases the inferred experimental ion temperatures and gives line of sight variations. 
\end{abstract}

\maketitle

\section*{\label{sec:level1}Introduction}

Fusion reactions between light elements may yield a clean source of nuclear energy, with few long-lived waste products. To achieve sufficient energy gain with a containable yield, the fuel must be compressed to much greater than solid density. This has been achieved using laser ablative compression, for example at the national ignition facility \cite{le2018fusion, PhysRevLett.114.255003, hurricane2014fuel}. Achievement of ignition would have far reaching applications in nuclear physics and experimental stockpile stewardship.

In the typical scheme, ignition occurs in a hot central region of the compressed fuel, known as the hot-spot. The burn wave then propagates into the colder and denser surrounding fuel. Sufficient fusion reactivity requires hydrogen temperature $T>4.5\,$keV. If the hot-spot areal density $\rho r>0.3\,$gcm$^{-2}$, fast alpha particles are mostly confined to the hot-spot and conduction losses are tolerable \cite{atzeni2004physics}. 

Previous kinetic numerical studies \cite{sherlock2009persistence, michta2010effects} found that particle collisions keep the fuel ion distributions close to Maxwellian, despite depletion of the faster ions due to fusion. The Maxwellian averaged fusion reactivity is accurate to better than 1\%. However, a more serious reactivity reduction occurs due to dense fuel boundaries \cite{peigney2014fokker, molvig2012knudsen, cohen2014one, albright2013revised, kagan2015self}. This is because the fast ions have a long mean free path and may be absorbed by the boundary layer, reducing yield \cite{molvig2012knudsen}. Kinetic effects during the implosion can also lead to deuterium tritium species separation \cite{inglebert2014species, PhysRevLett.122.035001}.

Another significant consideration is the incursion of the surrounding carbon ablator material into the hot fusion core \cite{ma2017role, PhysRevE.98.031201}, introducing further boundaries within the hot-spot. Bremsstrahlung radiation increases rapidly with ion atomic number, lowering the temperature and fusion yield in carbon mix regions. 

In this work, we adapt the Fokker-Planck method of reference \cite{larroche2012ion} to simulate the evolution of the hot-spot with localized regions of ablator mix. Increased electron bremsstrahlung radiation causes rapid cooling and leads to steep temperature gradients and non-local effects. The mix region Knudsen layer reduces the fusion reactivity. We find that radiative contraction of the mix region increases its density and therefore its radiative losses. Acting as a barrier to alpha particle transport, it reduces the effective hot-spot areal density. These effects make a localized mix spike more detrimental than uniformly distributed mix. The contraction also contributes to the broadening of the fusion neutron spectrum, leading to overestimates in the inferred ion temperature. 

 Experimental presence of ablator mix was confirmed with a corresponding increase of X-ray yield on lower fusion yield shots \cite{PhysRevLett.111.085004, PhysRevLett.111.045001, PhysRevLett.114.145004}. Ablator mix may occur due to diffusion near the time of stagnation \cite{rinderknecht2014first, rinderknecht2014kinetic} or target engineering defects exacerbated by the non-linear Rayleigh-Taylor instability \cite{hammel2011diagnosing, PhysRevLett.112.025002}. These asymmetries can lead to jets of carbon or tungsten dopant through the hot-spot. Localized spikes of mix have been confirmed with X-ray self emission images \cite{meezan2016indirect}, showing increased emission from the tent and fill tube positions. 

Three-dimensional radiation-hydrodynamic simulations also produced jets of ablator mix \cite{clark2016three, PhysRevLett.112.025002}. Radiative cooling produces steep temperature gradients around the mix region. The estimated hydrogen ion mean free path was $\lambda_\text{mfp}=50\,$nm and the temperature gradient scale-length was $5\,\mu$m, giving a Knudsen number $\lambda_\text{mfp} |\nabla T|/T$ of around $0.01$, requiring kinetic corrections to heat flow and fusion reactivity \cite{bell1981elecron, PhysRevLett.121.125001, molvig2012knudsen}. The development of the fast alpha particle spectrum may also be influenced by the mix region and its rapid evolution. As a result, although a fluid simulation can predict the existence of these mix regions, they incur additional hot-spot physics that can only be treated with a kinetic model.

At the ignition threshold with $T=5\,$keV and density $\rho = 100\,$gcm$^{-3}$, the Coulomb logarithm $\ln(\Lambda) = 0.5\ln(1+\lambda_\text{D}^2/b_\text{min}^2)\simeq 3$, where $b_\text{min}$ is the minimum impact parameter and $\lambda_\text{D}$ is the Debye length \cite{lee1984electron}. Particle collisions will affect the plasma evolution and can be modelled using the Fokker-Planck operator \cite{rosenbluth1957fokker}. Typical collision timescales are sub-picosecond, much less than the fuel stagnation timescale of $100\,$ps, so the electron and hydrogen ion distributions are expected to be near Maxwellian. This permits a simplified kinetic model allowing small deviations from Maxwellian distributions, allowing assessment of the kinetic corrections to fusion reactivity. 

In the first section of this paper, we describe the truncated ion kinetic model and implementation. The second section analyses a one-dimensional simulation and the additional insights from the kinetic model. The third section discusses the kinetic effect that reduces yield around the mix region. The fourth section introduces a two-dimensional simulation and the fifth section analyses it to show how the mix region broadens the neutron spectrum. The final section summarises the results.

\section*{Truncated Kinetic Model}

The hot-spot is modelled in one or two Cartesian spatial dimensions, with periodic boundaries. The physical model follows that of Keskinen \cite{PhysRevLett.103.055001}, in which the distribution functions $f_a$ (where $a$ labels the species) are decomposed in spherical coordinates in velocity space with a Cartesian tensor expansion. Since the collisions will cause $f_a$ to tend towards isotropy, the higher order terms will quickly decay and the expansion can be accurately truncated. The truncation is valid so long as the mean fluid velocity is less than the thermal velocity, which should be true after efficient conversion of the implosion energy to hot-spot internal energy at stagnation. This diffusive approximation yields
\begin{align}
f_a(t, \mathbf{x}, \mathbf{v}) = f_{0a}(t, \mathbf{x}, v) + \mathbf{\hat{v}.f}_{1a}(t, \mathbf{x}, v),\label{expansion}
\end{align}
where $v=|\mathbf{v}|$, $\mathbf{\hat{v}}=\mathbf{v}/v$ and $\mathbf{v}$ is the particle velocity. This decomposition decreases the dimensionality of the $f_a$ while retaining arbitrary anisotropic deviation in any Cartesian direction. The quantity $f_{0a}$ is the isotropic part of the distribution function and $\mathbf{f}_{1a}$ is the anisotropic part. Although the angular resolution of the anisotropy is reduced by this approximation, it still allows emergent kinetic phenomena such as Landau damping and flux limited heat flow.

The moments of the distribution functions give the number density $n_a$, fluid velocity $\mathbf{u}_a$, energy density $U_a$ and temperature $T_a$, for ion species $a$ of mass $m_a$ and charge $q_a$, via the integrals
\begin{align}
    n_a &= 4\pi\int_0^\infty f_{0a}v^2\,dv,\label{density}\\
    \mathbf{u}_a &= \frac{4\pi}{3n_a}\int_0^\infty \mathbf{f}_{1a}v^3\,dv,\label{velocity}\\
    U_a &= 2\pi m_a\int_0^\infty f_{0a}v^4\,dv,\\
    U_a &= \frac{1}{2}n_am_a|\mathbf{u}_a|^2 + \frac{3}{2}n_ak_\text{B}T_a.\label{energy}
\end{align}
The diffusive approximation is valid so long as the contribution of the first term in equation (\ref{energy}) is small compared to the second term. This rules out modelling of the implosion (as previously achieved with one-dimensional ion-kinetic codes \cite{taitano2018yield}) but does allow efficient two-dimensional modelling of the stagnated hot-spot. As such, simulations will be initialized in the time of peak fuel compression. This approximation also breaks down during the later stages of ignition and burn wave propagation, so the simulations will be curtailed once the approximation (\ref{expansion}) is no longer valid.

Inserting equation (\ref{expansion}) into the non-relativistic Vlasov-Fokker-Planck equation and averaging over angles leads to the kinetic equations for each ion species \cite{thomas2012review, tzoufras2011vlasov},
\begin{align}
\frac{\partial f_{0a}}{\partial t} +  \frac{v}{3}\mathbf{\nabla . f}_{1a} +\frac{q_a}{3m_av^2}\frac{\partial}{\partial v}(v^2\mathbf{E.f}_{1a})&= C_{0a} + F_a,\label{kinetic1}\\
\frac{\partial \mathbf{f}_{1a} }{\partial t} +     v\mathbf{\nabla} f_{0a} + \frac{q_a}{m_a}\mathbf{E}\frac{\partial f_{0a}}{\partial v} &= \mathbf{C}_{1a}.
\end{align}
In addition to these ion equations, the electrons are modelled as a fluid, with density and velocity found by assuming quasi-neutrality and electrostatic approximations, neglecting magnetic effects. This fluid approach is valid since the electron collision time is approximately $1\,$fs, much less than self-heating timescales. However, the ion collision timescales are not negligible, requiring the kinetic treatment. The quasi-neutral assumption gives the electric field from the electron pressure gradient $\mathbf{E} = -\nabla p_\text{e}/(n_\text{e}e)$, where $p_\text{e}=n_\text{e}k_\text{B}T_e$ and is found from equation (\ref{energy}). The electron energy density is evolved using a diffusion equation with Spitzer conductivity $\kappa_S$ and a flux limiter at 5\% of the free streaming value, 
 \begin{align}
     \frac{\partial U_\text{e}}{\partial t} + \mathbf{\nabla} .[ (p_\text{e}+U_\text{e})\mathbf{u_\text{e}} -\kappa_\text{S}\nabla T_\text{e}  ]= 0.
 \end{align}
The collision term $C_{0a}$ was given by the Rosenbluth-Fokker-Planck form \cite{rosenbluth1957fokker, tzoufras2011vlasov}, where the collision operator for each ion species $a$ is summed over collisions with all species $b$,
\begin{align}
    C_{0a}(\mathbf{x},v) &= \sum_b \frac{\Gamma_{ab}}{3v^2}\frac{\partial}{\partial v}\left(3\frac{m_a}{m_b}I_0^b +(I_2^b+J_{-1}^b)v\frac{\partial }{\partial v} \right)f_{0a},\\
    I^b_p(\mathbf{x},v) &=  \frac{4\pi}{v^p}\int_0^v  f_{0b}(\mathbf{x},w) w^{2+p} \,dw,\\
    J^b_p(\mathbf{x},v) &= \frac{4\pi}{v^p}\int_v^\infty f_{0b}(\mathbf{x},w) w^{2+p}\,dw,\\
    \Gamma_{ab} &= \frac{q_a^2q_b^2\ln(\Lambda)}{4\pi\epsilon_0^2m_a^2}.
\end{align}
The collision operator was implemented with the implicit, mass conserving method of Chang and Cooper \cite{chang1970practical}, allowing a time-step similar to the ion collision time. The $I$ and $J$ integrals for electrons were performed analytically by assuming a Maxwellian distribution, whereas the integrals were performed numerically for each ion species. Since the energy gained by ions must be lost by electrons, energy was conserved by updating $U_\text{e}$ accordingly. 

The collision operator $\mathbf{C}_{1a}$ is responsible for spreading of the particle direction. When the diffusive approximation (\ref{expansion}) is substituted in to the Fokker-Planck form given by equation (2) in ref. \cite{molvig2012knudsen}, it yields the velocity dependent Krook form \cite{PhysRevLett.103.055001, tzoufras2011vlasov},
\begin{align}
    \mathbf{C}_{1a} &= -\sum_b\frac{\Gamma_{ab}n_b}{v^3}\left(\mathbf{f}_{1a}-\mathbf{f}_{\text{M}a}\right), \\ \mathbf{f}_{\text{M}a} &= -\mathbf{u}_{\text{DT}}\frac{\partial f_{0a}}{\partial v}.
\end{align}
Although this is an approximation to the full Fokker-Planck form, it is most accurate for the supra-thermal ions responsible for the fusion reactivity reduction \cite{molvig2012knudsen}, a key kinetic effect. There is also a correction to the collision operator so that it more closely conserves momentum, tending towards $\mathbf{f}_{\text{M}a}$, the anisotropic part of a Maxwellian with fluid velocity equal to the deuterium-tritium fluid velocity. 

 The fusion term $F_a$ in equation (\ref{kinetic1}) used the Maxwellian fusion reactivity formula \cite{bosch1992improved}, giving the volumetric reaction rate $r$ as a function of temperature. There may be kinetic corrections to this formula, so the more accurate reactivity will be post-processed by integrating the full hydrogen distribution function from the simulation. The hydrogen ion density was reduced and the alpha particle distribution was increased using the local $r$ at each time-step. Alpha particles were initialized with energy of $3.5\,$MeV, with a small thermal spread. 

\begin{figure*}[t!]
  \centering
  \includegraphics{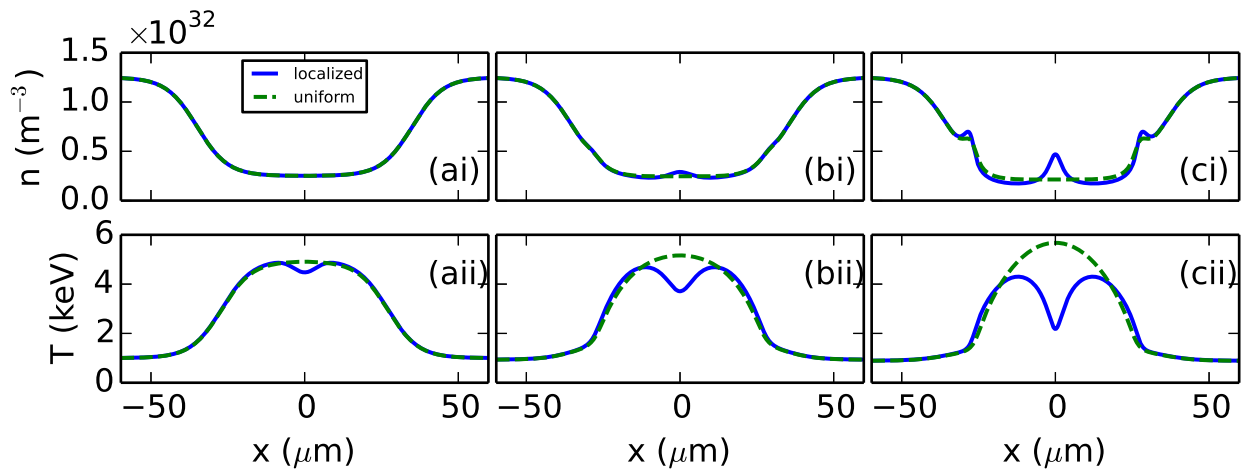}
  \caption{Results of two one-dimensional kinetic simulations, the first with carbon ablator mix in a Gaussian profile of waist $5\,\mu$m in the centre of the simulation domain and the second with the same carbon mass spread uniformly across the hot-spot. The (i) panels in the top row show the number density for the combined deuterium-tritium ion species at three separate times (a) $0\,$ps, (b) $20\,$ps and (c) $40\,$ps. The bottom row (ii) panels show the deuterium-tritium temperature. Data for other species are not shown. The carbon ions, assumed fully ionized, constituted $1.9\,$\% of the hot-spot mass and had peak number density $8\times10^{29}\,$m$^{-3}$ in the localized case and $10^{29}\,$m$^{-3}$ in the uniform case. }
  \label{f_simulation_nT}
\end{figure*}

The X-ray bremsstrahlung emission occurs due to collisions between electrons and ions and is a primary loss mechanism. The electron acceleration during collisions, and therefore the radiated power, increase with the ion atomic number $Z_a$. This means radiative losses will be greater within regions of carbon mix. The electron volumetric emission rate was calculated in reference \cite{atzeni2004physics}, as a function of electron temperature in eV and the ion number densities in cm$^{-3}$, 
\begin{align}
    W( T_\text{e}) = 1.69\times10^{-32}n_\text{e}\sqrt{T_\text{e}}\sum_an_aZ_a^2\,\text{Wcm}^{-3}.\label{rad}
\end{align}
 The sum is taken over all ion species, which are assumed fully ionized. Since equation (\ref{rad}) is only valid for weakly coupled plasma and not the dense fuel shell, the model imposes a smooth exponential cut-off of $W$ for regions of plasma with temperature below $1.5\,$keV.  The boundary conditions are open for the radiation and periodic for the plasma. The plasma is assumed to be optically thin, such that the radiation escapes.
 
 An additional effect that has not been considered is re-absorption of the bremsstrahlung radiation, which may occur in the central mix region and reduce its radiative cooling. The mix region spectrally averaged optical depth was estimated to approach 0.2 at the end of the simulation, meaning corrections due to re-absorption should be minimal.

\section*{One-Dimensional simulation}

Conditions were chosen close to the ignition threshold, with uniform initial pressure and a hot-spot with peak temperature $4.9\,$keV, fuel density $100\,$gcm$^{-3}$ and areal density $\rho r=0.35\,$gcm$^{-2}$. It was surrounded by a dense fuel shell with density five times that of the hot-spot. The one-dimensional Cartesian simulation domain had width $120\,\mu$m, 288 cells and periodic boundaries. The velocity grid consisted of 3000 points extending to $1.7\times 10^{7}\,$ms$^{-1}$. The deuterium and tritium ions were modelled as a single species with mass equal to $2.5$ atomic mass units. The addition of carbon and alpha particles made a total of three ion species. The time-step was $20\,$fs, sufficiently short to resolve the ion collision timescales. Note that although the boundaries are periodic, the dense plasma layer is thick enough to prevent transit of alpha particles through the boundary.
\begin{figure}[h!]
  \centering
  \includegraphics{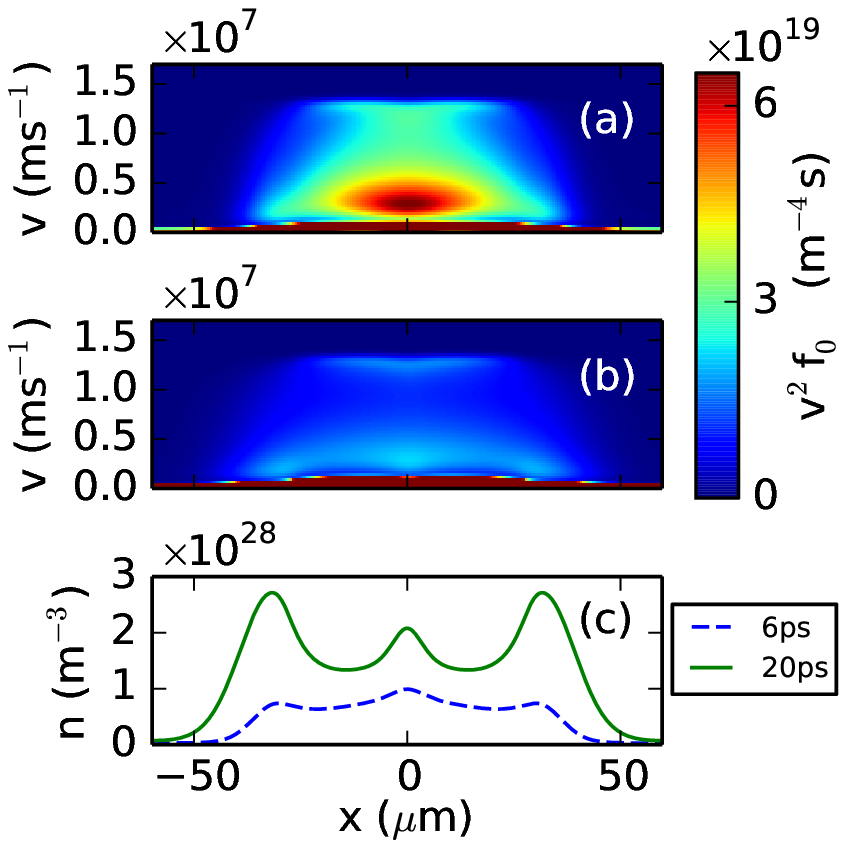}
  \caption{Distribution function $v^2f_0(x, v)$ of the supra-thermal alpha particles in the localized mix case, shown at times (a) $6\,$ps and (b) $20\,$ps. Fusion alpha particles are created with speed $1.3\times10^{7}\,$ms$^{-1}$. The thermalized alpha particles in the red stripe (exceeding the color scale) at the bottom of each plot have a thermal velocity $v_\text{th}\simeq \sqrt{k_\text{B}T_\text{e}/m_\alpha}=3\times10^5\,$ms$^{-1}$ and constitute the bulk of the alpha particle number density. The central mix region is more effective at slowing the alpha particles. Equation (\ref{density}) shows that the integral of these distribution functions along the $v$ axis gives the alpha particle number density, shown in panel (c) for both time-steps.}
  \label{fig:alphas}
\end{figure}
Fig. \ref{f_simulation_nT} shows the development of the plasma under the ion kinetic model. The upper panels show the number density and the lower panels show the temperature profiles of the deuterium-tritium species at the times $t=0, 20$ and $40\,$ps, for two separate simulations. The first simulation initialized the carbon ions in a localized Gaussian region of waist $5\,\mu$m and peak number density $8\times 10^{29}\,$m$^{-3}$, constituting $1.9\,$\% of the hot-spot mass. In the second simulation, the same mass of carbon was uniformly distributed across the hot-spot.
 
The carbon in the localized central mix region increases the average ion atomic number, which increases the bremsstrahlung radiation according to equation (\ref{rad}). This cools the electrons, which then cool the ions. The alpha particles transport energy from other parts of the hot-spot into the cooler mix region. This weakens the self heating and cools the whole hot-spot. Self-heating and ignition occurs in the uniform mix case, but not in the localized case.

Localized mix is more detrimental than the uniform mix because of its radiative cooling and contraction. The mix region reaches a much lower temperature within $40\,$ps. It contracts to several times the density of the surrounding hot-spot, as shown in the upper panels of Fig. \ref{f_simulation_nT}. This increases the stopping power for alpha particles. Furthermore, the energy deposited in the mix region is more efficiently radiated away and wasted, since the radiative rate scales as $n_\text{e}^2$. 

Once it cools and contracts, the mix region acts as an effective heat sink and barrier for the alpha particles, partially separating the two sides of the hot-spot and reducing the effective hot-spot areal density to below the threshold Lawson value. Although radiative losses are increased in the uniform case, the mix does not act as a barrier and there is no reduction of the hot-spot areal density. The uniform carbon mix acts purely to reduce the self heating rate across the hot-spot, whereas the localized mix region acts as a partial barrier and heat sink for alpha particles.

Fig. \ref{fig:alphas} shows the $f_0$ distribution function of the alpha particles across the hot-spot at two separate times, for the localized case. The distributions have been multiplied by the spherical velocity coordinates Jacobian $v^2$. The fusion alpha particles are created at a fixed speed, then slow down via collisions. The fuel temperature, and therefore the fusion reactivity, have decreased at the later time-step, so the supra-thermal phase space density is lower in Fig. \ref{fig:alphas}b than Fig. \ref{fig:alphas}a. The high phase space density for $v<10^6\,$ms$^{-1}$ is a result of the alpha particles that have slowed to form a thermal distribution with $T\simeq T_\text{e}$. The alpha particle stopping time is several picoseconds, so the thermal particles form the majority of the alpha particle population at the time-steps shown. 

It is clear from Fig. \ref{fig:alphas} that the central mix region is more effective at slowing the alpha particles, as there are more alpha particles at lower velocity within the mix region than outside of it. Since the radiative power scales as $n_\text{e}^2$, the alpha energy is more effectively radiated once the mix region becomes denser. This means the effects of the mix region become worse over time, an effect that is not observed for uniformly distributed mix.

Fig. \ref{fig:alphas}c shows the alpha particle number density for the two separate time-steps. There is a surplus of alpha particles within the dense mix region, even though the fusion reaction rate is lower here. This shows that a large fraction of alpha particles have travelled from other parts of the hot-spot into the mix region. They then slowed down and deposited their energy here, where it is radiated away and wasted. By efficiently slowing the alpha particles, the mix region acts in a similar way to the dense plasma shell. This effect becomes more pronounced as the mix region becomes denser at later times. These results show that localized carbon mix is more damaging and requires a greater ignition temperature than uniformly distributed carbon.
\begin{figure}[t]
  \centering
  \includegraphics{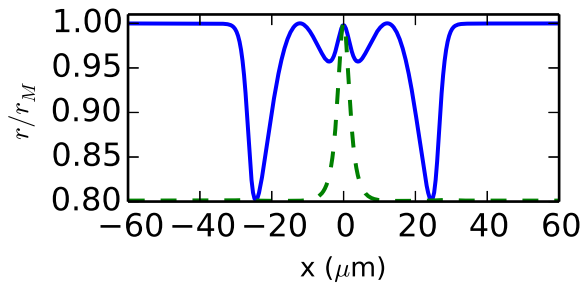}
  \caption{Fusion reactivity of the deuterium-tritium plasma from the one-dimensional simulation at $40\,$ps, calculated from a numerical integral of equation (\ref{reactivity}) using the simulated distribution function with localized carbon mix. The reactivity is normalized to the reactivity of a Maxwellian distribution with identical fluid moments. The normalized profile of the carbon mix is also overlaid with the dashed line.}
  \label{fig:reactivity}
\end{figure}
\begin{figure*}[ht!]
  \centering
  \includegraphics{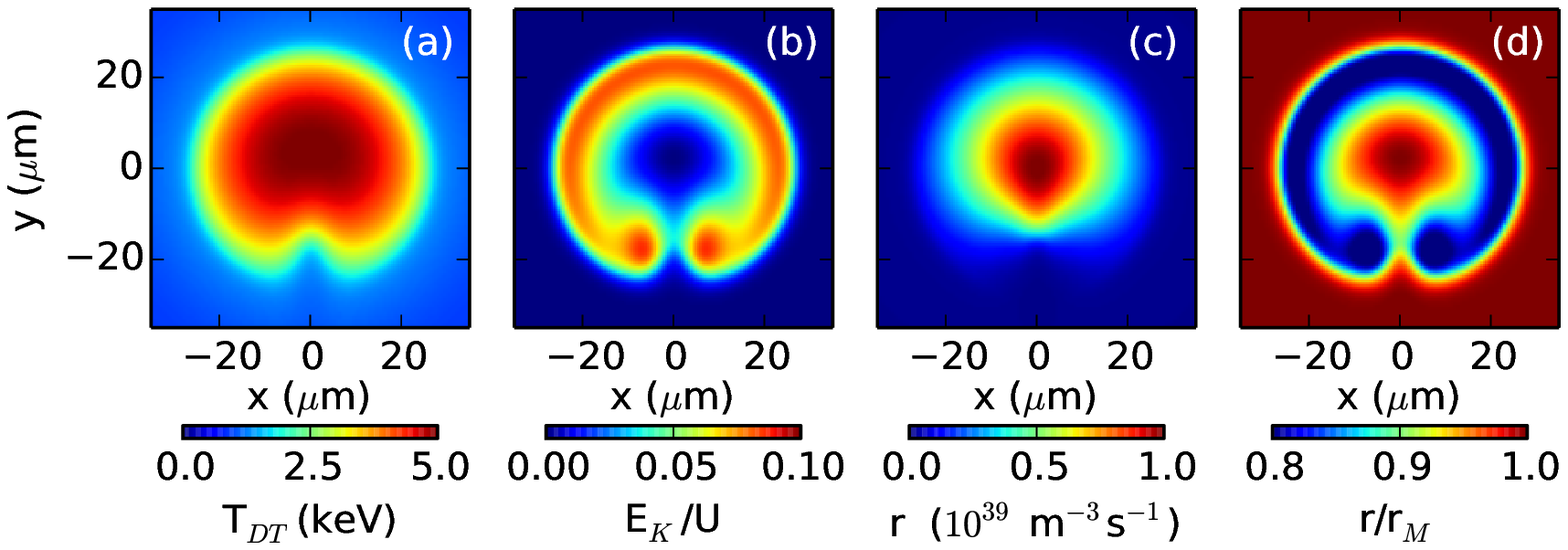}
  \caption{Results of the two-dimensional kinetic simulation after $40\,$ps, with carbon ablator mix localized in a Gaussian spike of waist $5\,\mu$m in the $x$ direction, waist $20\,\mu$m in the $y$ direction and centred at $y=-25\,\mu$m. The peak carbon number density was $8\times10^{29}\,$m$^{-3}$ and all other parameters are equivalent to Fig. \ref{f_simulation_nT}. (a) The temperature of the deuterium tritium species, showing the radiative cooling in the mix region. (b) The ratio of the deuterium-tritium fluid kinetic energy to the total energy, given by $mn|\mathbf{u}|^2/(2U)$. (c) The volumetric fusion reaction rate. (d) The ratio of the fusion reaction rate to that of a Maxwellian with equal density, temperature and fluid velocity.  }
  \label{2dreactivity}
\end{figure*}
\section*{Kinetic reduction of fusion reactivity}

Temperature gradients at the edge of the hot-spot are known to cause reductions to fusion reactivity through a purely kinetic effect \cite{molvig2012knudsen}. At the temperatures simulated, supra-thermal fuel ions are more likely to fuse, but also have a longer collisional mean free path, increasing the likelihood of their absorption by a nearby dense boundary layer. This can deplete the tail of the distribution and reduce fusion reactivity, even several thermal mean free paths from the boundary. 

The simulations in the present work show that this same mechanism also affects the edge of the mix region. Fig. \ref{fig:reactivity} shows the ratio of the simulated fusion rate with that of an equivalent Maxwellian distribution with equal fluid moments, calculated at the $40\,$ps time-step shown in Fig. \ref{f_simulation_nT}. The volumetric fusion reaction rate was calculated by numerically integrating the fusion cross section across the distribution function of the fuel ions in three velocity space dimensions,
\begin{align}
    r(\mathbf{x}) &=  \int\, d^3\mathbf{v'}\,d^3\mathbf{v}\,f_\text{D}(\mathbf{x, v'})f_\text{T}(\mathbf{x, v})\sigma(v_\text{r})v_\text{r}, \label{reactivity}
\end{align}
where $v_\text{r} = |\mathbf{v}-\mathbf{v'}|$ is the relative speed of the two interacting particles and $\sigma$ is the fusion cross section \cite{bosch1992improved}. 

The fusion rate is reduced in Knudsen layers at the edge of the hot-spot and near the mix region. The depletion of the tail of the distribution is strongest at the position of steep temperature gradients. The reduction is more severe at the edge of the hot-spot than around the mix region. 

The region of fusion reactivity reduction is wider than the carbon mix profile, shown by the dashed line, showing that the carbon has an extended effect beyond its boundaries. This is due to the long mean free path of the faster fuel ions. Within the mix region itself, the plasma is denser and more collisional and so the distribution is closer to Maxwellian.

\section*{Two-Dimensional simulation}

Localized mix regions may result from the Rayleigh-Taylor instability, resulting in an extended jet shape in two dimensions. The simulation was repeated in two Cartesian dimensions using equivalent parameters and a circular shaped hot-spot with radius $35\,\mu$m. The mix region was initialized using a Gaussian jet with waist $5\,\mu$m in the azimuthal direction and waist $20\,\mu$m in the radial direction. The centre of the Gaussian was offset in the $y$ direction by $25\,\mu$m from the centre of the simulation domain. The peak carbon ion number density was $8\times10^{29}\,$m$^{-3}$.

Fig. \ref{2dreactivity}a shows the deuterium-tritium temperature profile after $40\,$ps. The temperature is reduced in the mix region by radiative losses. This alters the shape of the hot-spot to become more elongated. 

Fig. \ref{2dreactivity}b shows the ratio of the deuterium-tritium kinetic fluid energy to the total energy, given by $mnu^2/(2U)$. This is a measure of the fluid flows in the hot-spot, and also indicates the applicability of the approximation given by equation (\ref{expansion}). Due to the radiative cooling around the edge of the hot-spot, there is a pressure gradient which causes fluid flow. There is also a flow of similar magnitude towards the mix region, which has steep pressure gradients due to the radiative cooling from the increased carbon concentration.

Due to its lower temperature, the fusion reaction rate $r$ is reduced in the mix region. Furthermore, inflow of the fuel to the mix region reduces the fuel density in regions around the mix spike. This means the reduction of fusion rate extends beyond the boundaries of the mix region. Fig. \ref{2dreactivity}c shows the fusion reaction rate $r$, as calculated using equation (\ref{reactivity}). Due to the decreased density and the kinetic effect, the shape of the hot-spot, as measured by the fusion burn profile, is more flattened than the shape measured by the temperature profile.

There may be experimental signatures from the localized carbon region, since the fusion neutron hot-spot image [similar to Fig. \ref{2dreactivity}c] is a considerably different shape to the hot-spot X-ray emission and temperature profile [Fig. \ref{2dreactivity}a]. This may allow increased diagnostic confidence when differentiating between uniform or localized mix. 

 Fig. \ref{2dreactivity}d shows the ratio of the fusion reactivity to that of an equivalent Maxwellian fuel distribution with equal fluid moments. The Knudsen layers reduce the reaction rate around the regions of high temperature gradients. The reduction is maximal around the edge of the hot-spot, reaching up to $20\,\%$. A similar reduction is also present at the edge of the mix region. The reduction is also more severe at the transverse edges of the jet, rather than at its tip.

The reactivity reduction of approximately $10\,\%$ will also be expected in regions between the multiple mix jets predicted by hydrodynamic simulations \cite{PhysRevLett.112.025002}, a volume comprising a large fraction of the hot-spot. The $10\,\%$ value obtained in these simulations is in approximate agreement with theoretical estimates \cite{molvig2012knudsen}.

\section*{Neutron Spectrum}

The contraction of the mix region may help to explain why the measured experimental ion temperature (from the width of the neutron spectrum) is anomalously high compared to that expected from fluid simulations and the measured fusion yield \cite{hurricane2014fuel, weber2015three}. This could be due to a significant fraction of the hot-spot energy residing in fluid kinetic energy rather than internal energy \cite{johnson2016indications, woo2018effects, murphy2014effect}. There is an intrinsic neutron spectrum width due to ion thermal motion, but fluid flows will also cause broadening. There is residual fluid motion due to incomplete shock stagnation. On top of this, the simulations here show that localized regions of ablator material in the hot-spot may radiatively cool, causing rapid contraction and additional fluid flows. Furthermore, the mix spike is asymmetric, leading to line of sight variations in the measured neutron spectrum.

\begin{figure}[t]
  \centering
  \includegraphics{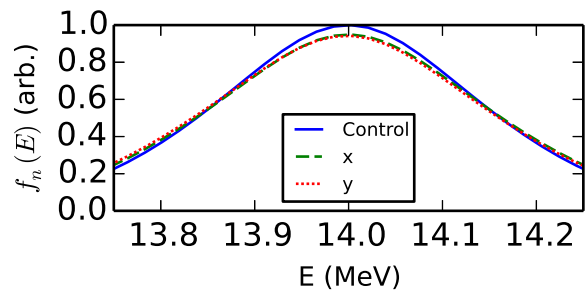}
  \caption{Synthetic normalized neutron spectra generated from the $40\,$ps time-step of the two-dimensional simulation shown in Fig. \ref{2dreactivity}. The solid line shows the case where the broadening from fluid motion is neglected, the dashed line shows the calculation with the broadening due to fluid motion for a line of sight in the $x$ direction, and the dotted line shows the same for the $y$ direction. The fitted temperatures are $4026\,$eV, $4464\,$eV and $4526\,$eV respectively. Neutron scattering was neglected. }
  \label{fig:neutron}
\end{figure}
To investigate this effect, we calculated the simulated DT fusion neutron spectra from the simulation shown in Fig. \ref{2dreactivity}, at the $t=40\,$ps time-step. There is a large inflow towards the cooler mix region. The neutron spectrum has a thermal broadening around the central energy $E_0\simeq 14\,$MeV, given by \cite{brysk1973fusion}
\begin{align}
    f_\text{n} &\propto  \int\,d^3\mathbf{x}\, r\exp\left(-\frac{(E-E_0)^2}{ k_\text{B}TE_0 }  \frac{m_\text{n}+m_\alpha}{4m_\text{n}}\right),\label{neutronspectrum}\\
    E(\mathbf{x},v) &= \frac{1}{2}m_\text{n}(v-\mathbf{u}_\text{DT}.\mathbf{\hat r})^2,
\end{align}
where $\mathbf{\hat r}$ is a unit vector along the line of sight, $r$ is the reaction rate defined in equation (\ref{reactivity}), $\mathbf{u}_\text{DT}$ is the fuel fluid velocity and $m_\text{n}$ is the neutron mass. The neutron spectrum around $v_0\simeq 5.2\times 10^7\,$ms$^{-1}$ is broadened by the ion thermal motion and shifted by the local ion fluid velocity, which has a peak value $|\mathbf{u}_\text{DT}|\simeq 2\times 10^5\,$ms$^{-1}$. The spatially integrated neutron spectrum from the whole fuel must be weighted by the fusion reaction rate $r$ at each point. Since $\mathbf{u}_\text{DT}.\mathbf{\hat r}$ is both positive and negative, the spatially integrated spectral shift equates to an additional spectral broadening. Equation (\ref{neutronspectrum}) assumes that the neutrons are not scattered by collisions as they escape. 

The calculation was performed to find the broadening for lines of sight in the $x$ and $y$ directions [Fig. \ref{fig:neutron}]. The numerical integration of equation (\ref{neutronspectrum}) for the simulation data at $t=40\,$ps was compared to the same integration with $\mathbf{u}_\text{DT}$ artificially set to zero. This negates the spectral shift from the fluid motion and leaves only the thermal broadening.

The spectra have an increased width as a result of the integrated fluid motion. This is a result of the motion at the edge of the hot-spot and is also partially due to the contraction of the mix region. There is a slight difference in broadening in the $x$ and $y$ directions, as well as a small shift of the spectrum central energy for the $y$ direction case. This is due to the contraction of the mix region in the $y$ direction. Since the effects from the hot-spot edge are isotropic, these differences between the $x$ and $y$ lines of sight are purely due to the mix region.

If the width of the spectra were to be used as an experimental ion temperature diagnostic, a fit of equation (\ref{neutronspectrum}) to the spectra yields $T = 4026\,$eV in the control case, $T = 4464\,$eV along the $x$ direction and $T = 4526\,$eV in the $y$ direction. Therefore the inflow of hot fuel into the mix region and dense fuel shell is enough to significantly broaden the fusion neutron distribution. The magnitude of this effect is similar to the known discrepancy at the National Ignition Facility \cite{hurricane2014fuel}. The variance in temperatures from different lines of sight are also consistent with experiments. Although the fluid flows towards the mix region have a greater magnitude in the $x$ direction than the $y$ direction, the $y$ neutron spectrum is broader because the fusion reaction rate and neutron emission is much greater at the tip of the mix spike than at its sides.

\section*{Summary}

We have used an ion kinetic model to compare the cases of uniformly distributed and localized carbon mix. The bremsstrahlung effects of localized mix were found to be more severe than the same carbon mass spread uniformly across the hot-spot. The localized mix region radiatively cools and contracts, increasing the radiative losses which scale as $n_\text{e}^2$. This contraction induces strong flows which broaden the fusion neutron spectrum, possibly affecting experimental ion temperature measurements. The cool, low density mix region acts as a barrier to alpha particles, separating different parts of the hot-spot and decreasing the effective hot-spot areal density. Kinetic depletion of the distribution tail around the mix region reduces the fusion reactivity below that of an equivalent Maxwellian distribution. Although all carbon mix will raise the ignition threshold, the present work indicates that localized jets of mix are especially damaging.

\begin{acknowledgments}
This work has been carried out within the framework of the EUROfusion Consortium and has received funding from the Euratom research and training programme 2014-2018 (CpfP-WP14-ER-01/CEA-12) and 2019- 2020 (ENR-IFE19.CCFE-01) under grant agreement no. 633053. Research presented in this article was supported by the Laboratory Directed Research and Development program of Los Alamos National Laboratory under project number 20180040DR. This work has also been funded by EPSRC grant numbers EP/L000237/1 and EP/R029148/1, and by STFC grant numbers ST/M007375/1 and ST/P002048/1. This work was supported by the European Research Council (InPairs ERC-2015-AdG grant no. 695088). This work used the ARCHER UK National Supercomputing Service (http://www.archer.ac.uk) and the LANL Grizzly cluster.
\end{acknowledgments}



\end{document}